\documentclass[10pt]{article}
\usepackage{graphicx}
\pdfoutput=1

\def\Title#1{\begin{center} {\Large #1 } \end{center}}
\def\Author#1{\begin{center}{ \sc #1} \end{center}}
\def\Address#1{\begin{center}{ \it #1} \end{center}}

\newcommand\pubblock{\rightline{\begin{tabular}{l} Proceedings of the Second Annual LHCP\\ \pubnumber\\
         \pubdate  \end{tabular}}}

\newenvironment{Abstract}{\begin{quotation} \begin{center} 
             \large ABSTRACT \end{center}\bigskip 
      \begin{center}\begin{large}}{\end{large}\end{center} \end{quotation}}

\newenvironment{Presented}{\begin{quotation} \begin{center} 
             PRESENTED AT\end{center}\bigskip 
      \begin{center}\begin{large}}{\end{large}\end{center} \end{quotation}}

\def\beq{\begin{equation}}
\def\eeq#1{\label{#1}\end{equation}}
\def\eeqn{\end{equation}}

\def\beqa{\begin{eqnarray}}
\def\eeqa#1{\label{#1}\end{eqnarray}}
\def\eeqan{\end{eqnarray}}

\let\bar=\overbar

\def\Dslash{\not{\hbox{\kern-4pt $D$}}}
\def\dslash{\not{\hbox{\kern-2pt $\del$}}}

\def\msb{{\bar{\ssstyle M \kern -1pt S}}}

\usepackage{color}

\newcommand\pubnumber{ ATL-PHYS-PROC-2014-102 }

\newcommand\pubdate{\today}

\def\affiliation{
On behalf of the ATLAS Collaboration, \\
Laboratoire de Physique Nucl\'eaire et de Hautes Energies (LPNHE)\\
UPMC, Universit\'e Paris-Diderot, CNRS/IN2P3, Paris, France }

\def\support{\footnote{Work supported by the Polish France collaboration IN2P3-COPIN 10-140}}

\begin{document}

\large
\begin{titlepage}
\pubblock

\vfill
\Title{ATLAS results on top properties}
\vfill

\Author{ Fr\'ed\'eric Derue \support }
\Address{\affiliation}
\vfill
\begin{Abstract}

Recent measurements of top quark properties using $t{\bar t}$ events produced in proton-proton 
collisions at the Large Hadron Collider with centre of mass energies of 7 and 8 TeV and detected
by the ATLAS experiment are presented. These results include top quark mass,
the top and anti-top mass difference, the electric charge, the top quark polarization and 
spin correlation, the $t{\bar t}$ charge asymmetry and the search for 
flavour changing neutral currents.

\end{Abstract}
\vfill

\begin{Presented}
The Second Annual Conference\\
 on Large Hadron Collider Physics \\
Columbia University, New York, U.S.A \\ 
June 2-7, 2014
\end{Presented}
\vfill
\end{titlepage}
\def\thefootnote{\fnsymbol{footnote}}
\setcounter{footnote}{0}
%

\normalsize 


\section{Introduction}

The datasets collected by the Large Hadron Collider (LHC~\cite{LHC}) in 2011 and 2012 
open a precision era in the study of the properties of the top quark. 
This quark is the most massive elementary particle known to date, decays almost
exclusively through a single decay mode ($t\rightarrow Wb$) and, due to its extremely short 
lifetime of $\sim5\times 10^{-24}$ s, decays before hadronization, allowing its spin information 
to be passed to the decay products. 
Such unique characteristics provide not only an excellent way of testing the
Standard Model (SM), but also an important window to physics beyond it.
A large variety of top quark properties can be studied for top quark pair production, either in
production, decay, for the whole $t{\bar t}$ system or single top quarks. 
These proceedings focus on the measurement of its mass, the electric charge, 
the top quark polarization and spin correlation, the $t{\bar t}$ charge asymmetry 
and the search for flavour changing neutral currents.

All of the measurements discussed herein are the results of the analysis of 4.7 ${\rm fb^{-1}}$ 
of 7 TeV and 20.3~${\rm fb^{-1}}$ of 8 TeV 
proton-proton collisions collected by the ATLAS detector~\cite{ATLAS}. 
%
\section{Mass measurements}
The most defining property of a particle is probably the value of its mass.
In the case of the top quark, its mass plays an important role in particle physics, 
mainly because of its large coupling to the Higgs boson. 
the top quark mass ($m_{{\rm top}}$) plays indeed an important role in electroweak radiative 
corrections which yield a quadratic dependence of the W boson mass ($m_{{\rm W}}$) on $m_{{\rm top}}$, 
while $m_{{\rm W}}$ depends only logarithmically on the Higgs boson mass ($m_{{\rm H}}$).
Therefore, the precise measurement of $m_{{\rm top}}$, $m_{{\rm W}}$ and $m_{{\rm H}}$ 
provides an important consistency test of the Standard Model.
Besides, calculations at next-to-next to leading order, assuming the SM to be valid up
tho the highest energies, suggest that the Higgs potential might be unstable at high energies close
to the GUT scale ($10^{16}$ GeV), with this feature strongly dependent on the value of the top pole
mass~\cite{mtwh}. 
Setting aside the theoretical implications, from an experimental perspective this gives a
strong motivation for a precise determination of this parameter.
Usually, the top mass is measured as an invariant mass of its decay products, with corrections
applied using Monte Carlo simulations. 
Unfortunately, beyond the leading order its value depends on the renormalization scheme, 
which is not well-defined in current Monte Carlo generators. 
This implies an ambiguous definition when the uncertainty is below 1 GeV and close 
to the QCD scale ($\Lambda_{QCD}\sim 0.5$ GeV).

ATLAS performed several measurements of the top mass. 
The most precise one is obtained
with a three-dimensional fit to constrain the largest systematics~\cite{ATLASCONF2013046}. 
The first step is the full reconstruction of the $t{\bar t}$ decay kinematics 
on an event-by-event basis. 
This is accomplished in the lepton+jets channel with a maximum likelihood approach, 
whose inputs are the four jets, the charged lepton and the missing transverse energy. 
The kinematic fit extracts the top mass using the jets permutation with the highest likelihood. 
This information is used to reconstruct the mass of the hadronically-decaying $W$ boson, 
which depends on the jet energy scale, and a parameter called $R_{{\rm lb}}$ defined as the ratio 
between the transverse momenta of $b$-jets and light-jets in the hadronic decay of the top quark. 
This quantity, being a ratio, is independent of the jet energy scale by construction. 
A large number of templates are generated for different top mass hypotheses in the experimentally 
allowed range, and for different energy-scale factors of light jets (JSF) and $b$-jets (bJSF). 
An unbinned 3-dimensional fit extracts the most probable values of the top mass and the energy 
scale factors.
These factors turn out to be close to unity. 
The total uncertainty on the final result is less than 1\% :
\begin{eqnarray}
m_{{\rm top}}  &=& 172.31 \pm 0.75 {\rm (stat+JSF+bJSF)} \pm 1.35 {\rm (syst)} \,  {\rm GeV} \ .
\end{eqnarray}
The main uncertainties are the $b$-tagging efficiency, the jet energy scale and the 
statistical component of the $b$-jet energy scale, which will be reduced with more data. 
%
\begin{figure}[htb!]
\centering
\includegraphics[height=1.9in]{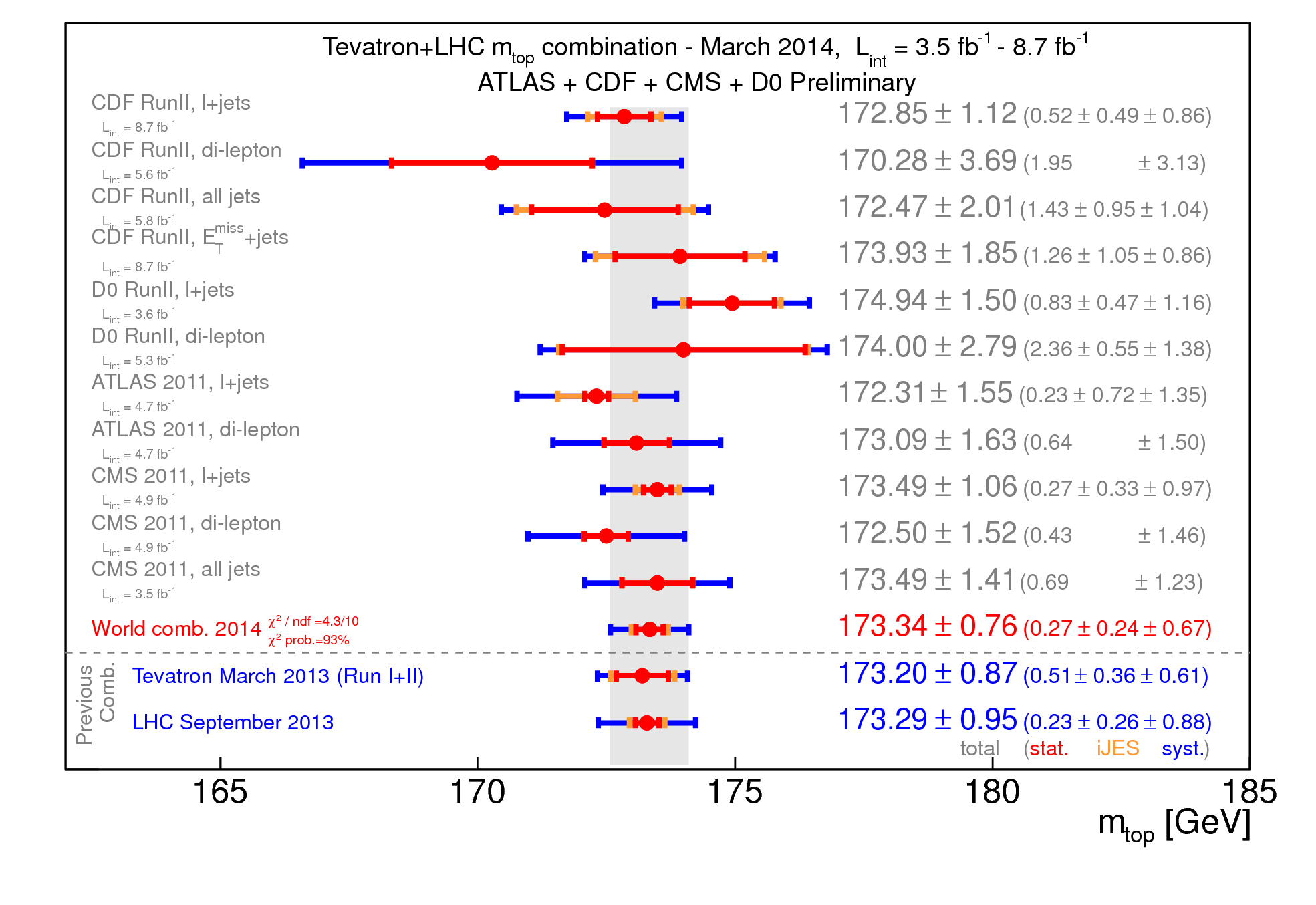}
\caption{Input $m_{{\rm top}}$ measurements and result of their combination from the Tevatron and LHC. For each measurement, the total uncertainty, the statistical and the iJES (Jet Energy Scale) contributions (when applicable), as well as the sum of the remaining uncertainties are reported separately. The iJES contribution is statistical in nature and applies only to analyses performing in situ (top quark pair based) jet energy calibration procedures. The grey vertical band reflects the total uncertainty on the combined $m_{{\rm top}}$ value.}
\label{fig:topmass}
\end{figure}
%
%
%

The ATLAS and CMS measurements of the top quark mass in different topologies were
combined with the results obtained by the CDF and D0 experiments at the Tevatron, as
summarized in Fig.~\ref{fig:topmass}. 
This combination was done with the Best Linear Unbiased Estimate
(BLUE) method and the individual measurements were found to be in good agreement, leading
to a $\chi^2$/ndf of 4.3/10. The current efforts by the different collaborations to harmonise the
treatment of the systematic uncertainties should be noted. The world combination result has a
total relative uncertainty of 0.4\%, dominated by the systematic 
uncertainties~\cite{masscombination}.

ATLAS is also measuring an eventual difference between the top and anti-top masses
using 4.7 ${\rm fb^{-1}}$ of data at 7 TeV~\cite{massdiff}. 
The measurement is $m_{t}-m_{\bar t} = 0.67 \pm 0.61 {\rm (stat)} \pm 0.41 {\rm (syst)}$ GeV, 
with the dominant sources of systematic uncertainties being the choice of the $b$ quark 
fragmentation model (0.34 GeV) and the different calorimeter response to $b$ and ${\bar b}$ 
initiated jets (0.08 GeV).

\section{Electric charge}
In the SM, the top quark is an up-type quark so its charge is $+2/3$ 
of the electron charge. 
Shortly after the discovery of the top quark, an exotic alternative to the SM
was proposed in which a top-like particle appears with charge $-4/3$. 
The analysis strategy adopted by ATLAS~\cite{eleccharge} relies on the fact that 
in the SM the product of the charges of the lepton and that of the $b$-quark 
is always negative, while in the exotic model it is always positive.
The sign of the top quark charge is transferred to the $W$ boson and then to the charged lepton, 
and it is quite easy to measure this quantity experimentally. 
This is not true for the $b$-quark, because the charge of a jet is not a well-defined quantity. 
Experimentally the weighted sum of the track charges is used as proxy for this quantity.

A harsh set of cuts is applied to the collected events to isolate a very pure sample. 
The observable is then compared against the two models. 
The resulting top charge is $0.64\pm 0.02 {\rm (stat)} \pm 0.08 {\rm (syst)}$, 
compatible with $+2/3$,  as predicted by the SM. 
The exotic model is now excluded by more than 8 standard deviations. 
%
\section{Top spin related measurements}
In the SM, $t{\bar t}$ pairs are produced essentially unpolarized at the LHC but the 
correlation of the spin orientation of the top and the anti-top quark is predicted to be non-zero. 
New physics models beyond the SM can change the spin correlation of the top and the 
anti-top quark by either changing the spin of the daughter particles of the top and anti-top 
quarks, or by changing the production mechanism of the $t{\bar t}$ pair.

The spin correlation of $t{\bar t}$ pairs can be extracted by analysing the angular
distributions of the top quark decay products.
The differential distribution of the decay width, 
$d\Gamma/d|cos(\theta_{\pm})|$, is proportional to the cosine of the angle, $\theta_{\pm}$, 
between the positively (negatively) charged lepton from the top (anti-top) quark decay and 
the top (anti-top) quark spin quantization axis in the top (anti-top) quark rest frame.
The spin correlation is expressed by the double differential cross-section :
\begin{equation}
\frac{1}{\sigma} \frac{d^{2}\sigma}{d[\cos(\theta_{+})][\cos(\theta_{-})] }
 = \frac{1}{4} \left( 1 + \alpha_{+}P_{+}\cos(\theta_{+}) + \alpha_{-}P_{-}\cos(\theta_{-})
+A\alpha_{+}\alpha_{-}\cos(\theta_{+})\cos(\theta_{-}) \right) \ ,
\label{eq:spincorr}
\end{equation}
where $A$ is the asymmetry built from the ratio between the number of events where the top quark 
and anti-top quark spins are parallel and the number of events where they are anti-parallel,
$\alpha$ represents the spin analysing power and $P_{\mp}$ is the polarisation of the
top/anti-top quark.

Since $\alpha\approx 1$ at leading order for charged leptons, leptons are taken as the final 
state particles for the following measurements. 
As spin quantization axis, the helicity basis of the top quark is used since it is well defined 
and provides predictions from theory.
%
\subsection{Top quark polarization}
The measurement of the top quark polarization in $t{\bar t}$ events is performed in the lepton+jets
and dilepton channels~\cite{toppolarization}. 
To extract the polarization from the $\cos\theta$ distribution, templates with assumed 
polarizations of $\alpha P=\pm 0.3$ were created. 
Two different mechanisms were considered to account for CP conservation (CPC) as in the SM 
and CP violating (CPV) $t{\bar t}$ production.
For the CP conserving case, the top quarks are polarized in the same way whereas for the CP
violating case, the top quarks have opposite polarizations. 
This implies that for one of the $\cos\theta$ distributions (either top or antitop indicated 
by the lepton charge), the templates for CPC and CPV are swapped. 
With the templates, a binned likelihood fit is performed to extract the fraction of positive 
polarization $f$ from the $\cos\theta$ distributions. 
Together with the polarization, the $t{\bar t}$ cross section is fitted to reduce uncertainties 
coming from the normalization of the processes. 
The background normalization is not fitted during the procedure, but estimated from 
Monte Carlo (MC) and data driven techniques
The combination was done by multiplying the likelihoods of each channel. 
All the results are in good agreement with the SM within the uncertainties. 
The combined result of lepton+jets and dilepton channels for the two production mechanisms is then :
\begin{eqnarray}
\alpha_{l}C_{{\rm CPC}} &=& -0.035 \pm 0.014 {\rm (stat)} \pm 0.037 {\rm (syst)} \\
\alpha_{l}C_{{\rm CPV}} &=& -0.020 \pm 0.016 {\rm (stat)}^{+0.013}_{-0.017} {\rm (syst)} \ . \nonumber 
\end{eqnarray}
Dominant systematic uncertainties come from modelling of the signal MC and jet energy 
related quantities.
\subsection{Top quark spin correlation}
Top quark spin correlation has already been observed at
ATLAS at a significance of 5.1$\sigma$ using 2.1 ${\rm fb^{-1}}$ of data~\cite{spincorr21}.
The full dataset has been analysed to check the compatibility
with the SM spin correlation strength prediction~\cite{spincorrpred}.
The spin correlation strength $A$ gives the
difference between like-helicity and unlike-helicity top quarks and is defined as :
\begin{equation}
A = \frac{N_{\Uparrow\Uparrow}+N_{\Downarrow\Downarrow}-N_{\Uparrow\Downarrow}-N_{\Downarrow\Uparrow}}
{N_{\Uparrow\Uparrow}+N_{\Downarrow\Downarrow}+N_{\Uparrow\Downarrow}+N_{\Downarrow\Uparrow}} \ .
\end{equation}
In order to extract the spin correlation strength from the selected data, four different 
observables are investigated, which are different linear combinations of components 
in the spin density matrix of $t{\bar t}$ production: the azimuthal difference $\Delta\phi$ 
of the charged lepton momentum directions in the laboratory frame, the ratio of the squares 
of matrix elements for top quark pair production and decay from the fusion of like-helicity gluons 
with and without spin correlation at leading order (S-Ratio), and the product of $\cos(\theta_+)$ 
and $\cos(\theta_-)$\footnote{The subscripts "+" and "-" refer to the lepton’s charge.} 
in both the helicity basis and the so called maximal
basis, which maximizes the value of the spin correlation strength.
No prediction from the SM exists for the maximal basis, so the value from MC simulation 
$A_{{\rm max}}~=~0.44$ with MC@NLO+Herwig~\cite{MCatNLO} was taken.
The SM prediction for the helicity basis is $A_{{\rm hel}}~=~0.31$ at 7 TeV.
A template fit is then performed for all four observables using templates with SM spin correlation
and without it, both generated with MC@NLO. 
The result is expressed in the parameter $f_{{\rm SM}}$ with $f_{{\rm SM}}=1$ being the SM spin correlation 
case and $f_{{\rm SM}}=0$ being the uncorrelated case and is shown in Fig.~\ref{fig:chargeasymcomb} 
(left part) for the four different observables. 
To get the measured spin correlation strength, the $f_{{\rm SM}}$ obtained from the fit has to be 
multiplied with the SM prediction. 
The dominant systematic uncertainties come from $t{\bar t}$ modelling.
\section{$t{\bar t}$ charge asymmetry}
\begin{figure}[htb!]
\centering
\includegraphics[height=1.7in]{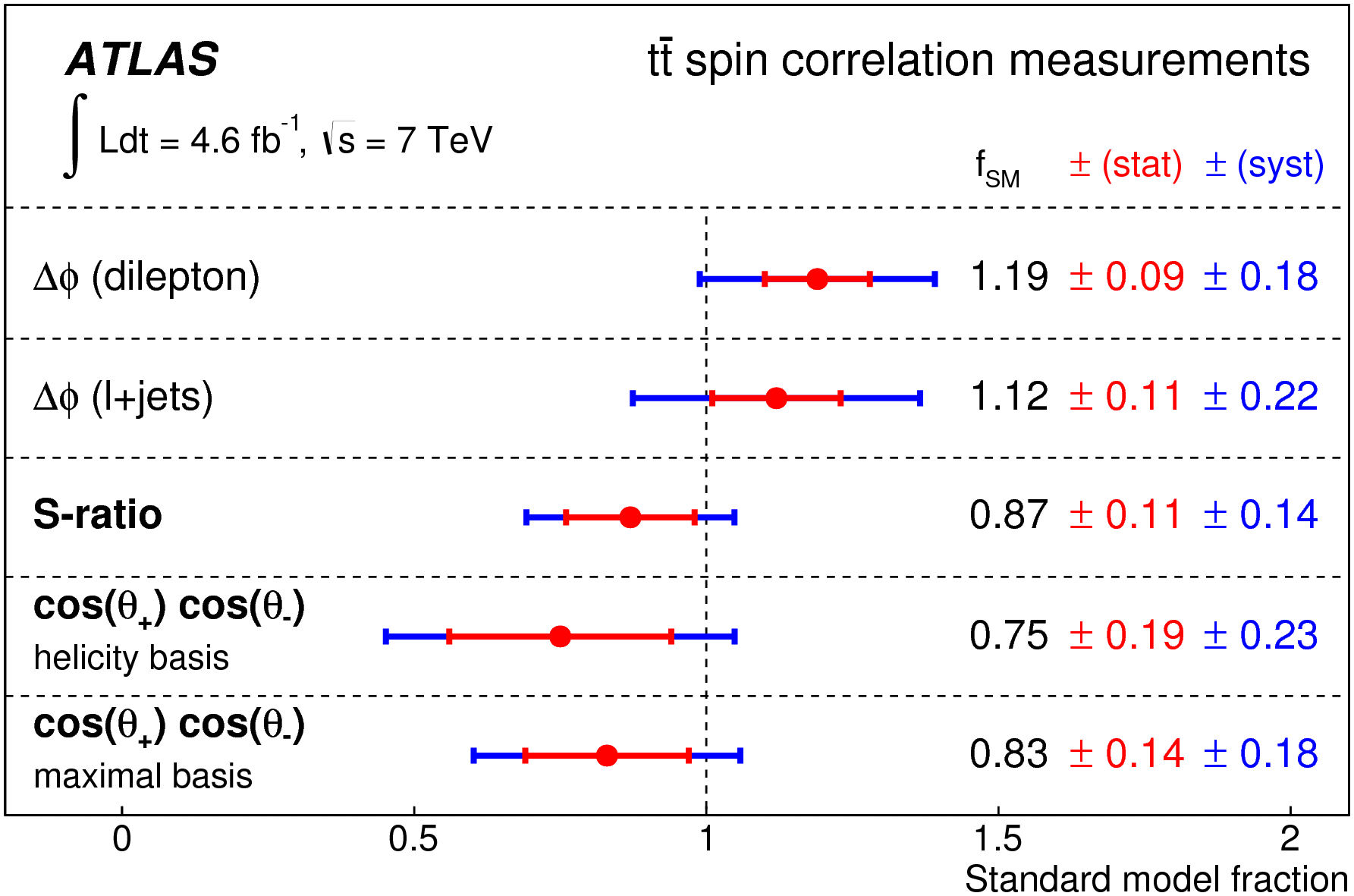}
\includegraphics[height=1.7in]{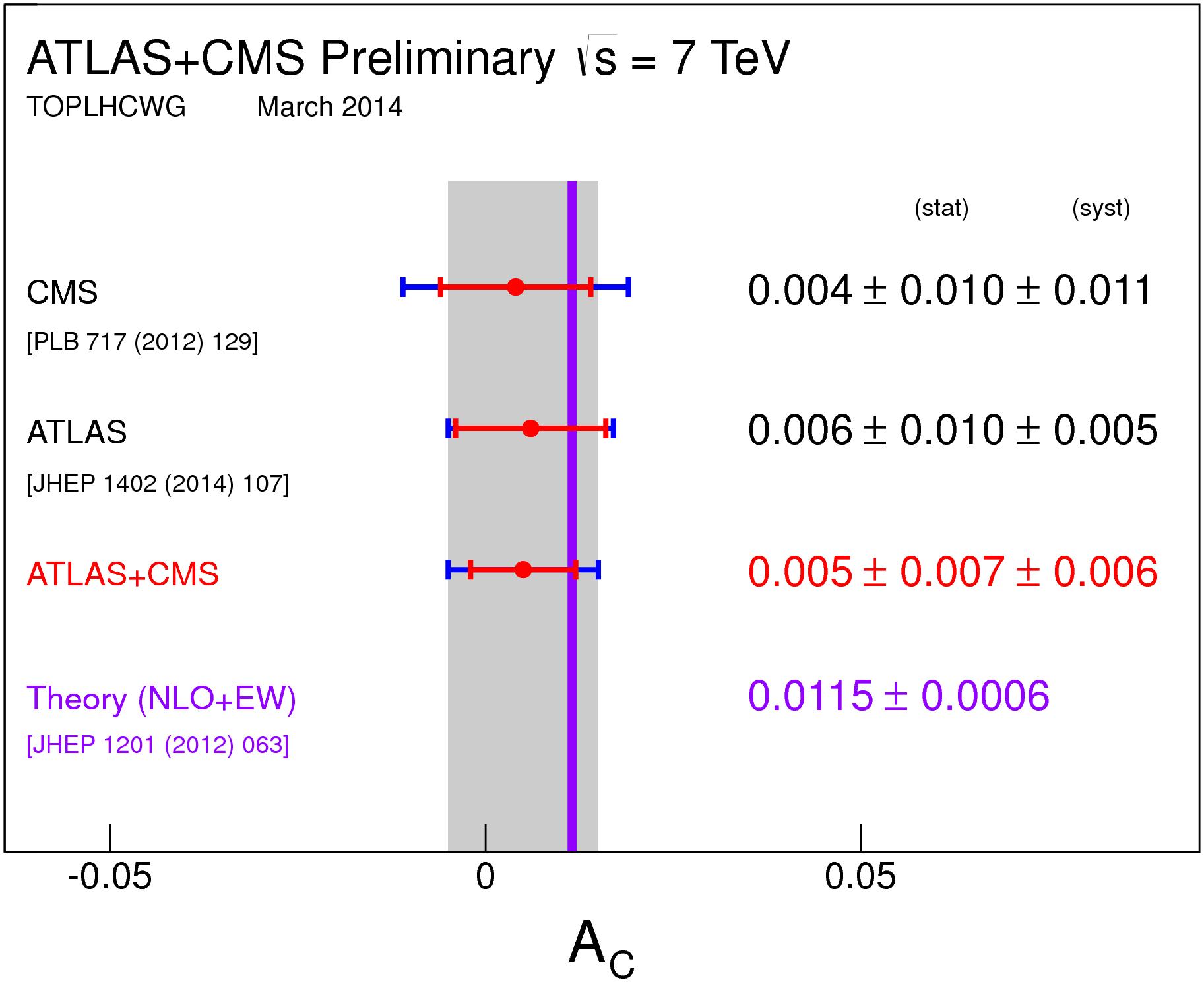}
\caption{Left~: summary of the measurements of the fraction of $t$ anti-$t$ events corresponding to the SM spin correlation hypothesis, $f_{{\rm SM}}$, in the dilepton final state, using four spin correlation observables sensitive to different properties of the production mechanism, and in the single-lepton final state. Dashed vertical line at $f_{{\rm SM}}$ = 1 indicates the SM prediction. The inner, red error bars indicate statistical uncertainties, the outer, blue error bars indicate the contribution of the systematic uncertainties to the total uncertainties. Right~: summary of the single measurements and the LHC combination of the $t{\bar t}$ charge asymmetry compared to the theory prediction (calculated at NLO including electroweak corrections). The inner red error bars indicate the statistical uncertainty, the blue outer error bars indicate the total uncertainty. The grey band illustrates the total uncertainty of the combined result.}
\label{fig:chargeasymcomb}
\end{figure}

The $t{\bar t}$ production is predicted by the SM to be symmetric at leading order (LO)
under charge conjugation. 
At NLO, however, for the $q{\bar q}$ and $qg$ production modes, there is a
small preference to produce the $t{\bar t}$ in the direction of the incoming quark 
(anti-quark).
It is therefore useful to define the following charge asymmetry,
$A_{c} = \frac{(\Delta|y|>0)-(\Delta|y|<0)}{(\Delta|y|>0)+(\Delta|y|<0)}$ , 
%
with $\Delta|y|=|y_{t}|-|y_{\bar t}|$, making use of the difference of absolute rapidities 
$|y_{t}|$ and $|y_{\bar t}|$ of top and antitop quarks.
At NLO in the SM it is expected to be $0.0115\pm 0.0006$ for $pp$ collisions at 7~TeV. 
In the lepton+jets channels, ATLAS measured the charge asymmetry in
$t{\bar t}$ events using 4.7~${\rm fb^{-1}}$ of data.
The collaboration unfolded the reconstructed $\Delta|y|$ distribution to parton level
and measured $0.006\pm 0.010 {\rm (stat)} \pm 0.005 {\rm (syst)}$\cite{chargeasym}. 
Furthermore, a combination~\cite{chargeasymcomb} with the results of the CMS experiment 
has been done and is presented in Fig.~\ref{fig:chargeasymcomb} (right part).
\section{Flavour Changing Neutral Currents}
In the SM, the top quark decays to a $W$ boson and a $b$-quark with a branching
ratio very close to unity. 
Non-standard couplings can affect the way a top quark can be produced or decays.
In the SM, these couplings do not exist at tree level, and are GIM-suppressed 
at higher orders

The $qg\rightarrow t$ analysis~\cite{FCNC1}, is performed on data collected in 2012, corresponding 
to an integrated luminosity of 14.2 ${\rm fb^{-1}}$ at 8~TeV.
FCNC $t\rightarrow qg$/$qg\rightarrow t$ is searched for in the production of single top-quark 
events.
No evidence of FCNC single top-quark production is found and the upper limit at 95\% CL on the 
production cross section is 2.5 pb. 
Using the NLO predictions for the FCNC single top-quark production cross-section and assuming
BR($t\rightarrow Wb$)= 1, the measured upper limit on the production cross-section is converted 
into limits on the coupling constants which in turn can be converted into limits on the branching
fractions: BR($t\rightarrow ug$)$<3.1\times 10^{-5}$ and BR($t\rightarrow cg$)$<1.6\times 10^{-4}$.
\begin{figure}[htb!]
\centering
\includegraphics[height=1.7in]{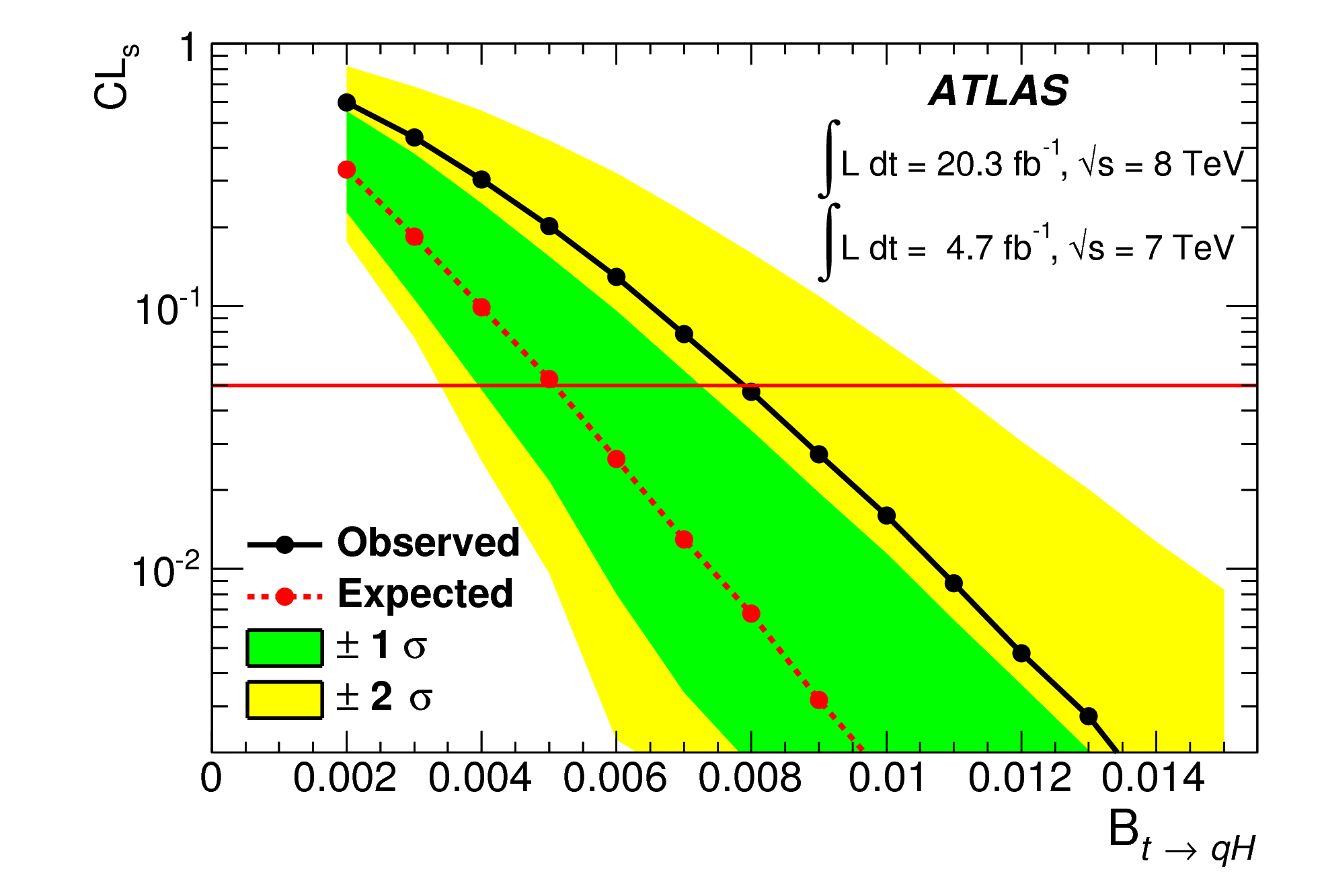}
\caption{Evolution of CLs as a function of the branching fraction B of the $t$ to $qH$ decay for the observation of a signal at 125.5 GeV (solid line) and the expectation in the absence of signal (dashed line). The 1 and 2$\sigma$ uncertainty bands around the expected curve are also shown.}
\label{fig:fcnc}
\end{figure}

A search is performed for FCNC in the decay of a top quark to an up-type (c,u) quark and a Higgs 
boson, where the Higgs boson decays to two photons~\cite{FCNC2}. 
The proton-proton collision data set used corresponds to 20.3 ${\rm fb^{-1}}$ at 8~TeV.
Top quark pair events are searched for in which one top quark decays to $qH$ 
and the other decays to $bW$. 
Both the hadronic and the leptonic decay modes of the $W$ boson are used. 
No significant signal is observed (cf. Fig.~\ref{fig:fcnc}) 
and an upper limit is set on the $t\rightarrow qH$
branching ratio of 0.79\% at the 95\% confidence level.
\section{Conclusions}
Thanks to the outstanding performance of the LHC, the experimental physics of the top quark
entered the precision era, with most measurements already dominated by systematic uncertainties.
The mass of the top quark is now known with a precision below 1 GeV. 
A big variety of top quark properties in top quark pair production and decay have been measured 
and presented.
All results are in good agreement with the SM and no sign of
new physics has been found so far.
 


\begin{thebibliography}{99}


\bibitem{LHC} 
  Lyndon Evans and Philip Bryant, 
  JINST 3 (2008) S08001

\bibitem{ATLAS} 
  ATLAS Collaboration,
  JINST 3 (2008) S08003

\bibitem{mtwh}
  J. Elias-Miro et al.,
  Phys.\ Lett.\ B {\bf 709} (2012) 222-228

\bibitem{ATLASCONF2013046} 
  ATLAS Collaboration,
  ATLAS-CONF-2013-046, https://cds.cern.ch/record/1547327.

\bibitem{masscombination} 
  ATLAS Collaboration,
  ATLAS-CONF-2014-008, CDF-NOTE-11071, CMS-PAS-TOP-13-014, D0-NOTE-6416 [arXiv:1403.4427 [hep-ex]]

\bibitem{massdiff} 
  ATLAS Collaboration,
  Phys.\ Lett.\ B {\bf 728C}, 363-379 (2014) 

\bibitem{eleccharge} 
  ATLAS Collaboration,
  JHEP11 (2013) 031 

\bibitem{spincorr21} 
  ATLAS Collaboration,
   Phys.\ Rev.\ Lett.\ 108 (2012) 212001

\bibitem{spincorrpred} 
  ATLAS Collaboration,
  subm. to PRD [arXiv:1407.4314 [hep-ex]]

\bibitem{MCatNLO} 
  S. Frixione, F. Stoeckli, P. Torrielli, B. R. Webber and C. D. White,
  arXiv:1010.0819

\bibitem{toppolarization} 
  ATLAS Collaboration,
   Phys.\ Rev.\ Lett.\ 111 (2013) 23, 232002

\bibitem{neutrino} 
 B. Abbott et al. [D0 Collaboration], Phys.\ Rev.\ Lett.\ 80 (1998) 2063

\bibitem{chargeasym} 
  ATLAS Collaboration,
  JHEP02 (2014) 107 

\bibitem{chargeasymcomb} 
  ATLAS and CMS Collaborations,
  ATLAS-CONF-2014-012, CMS-PAS-TOP-14-006, \\https://cds.cern.ch/record/1670535

\bibitem{FCNC1} 
  ATLAS Collaboration,
  ATLAS-CONF-2013-063, https://cds.cern.ch/record/1562777

\bibitem{FCNC2} 
  ATLAS Collaboration,
  JHEP06 (2014) 008

\end{thebibliography}
\end{document}